\documentclass[11pt]{article}
\usepackage{moriond,epsfig}

\def\Journal#1#2#3#4{{#1} {\bf #2}, #3 (#4)}

\def\NIMA{{\em Nucl. Instrum. Methods} A}

\def\PRL{\em Phys. Rev. Lett.}

\def\be{\begin{equation}}
\def\ee{\end{equation}}
\def\bea{\begin{eqnarray}}
\def\eea{\end{eqnarray}}

\begin{document}
\vspace*{3.4cm}
\title{THE TOTEM EXPERIMENT AT THE LHC}

\author{G. LATINO \\
(on behalf of the TOTEM Collaboration)}

\address{INFN Pisa $\&$ University of Siena, Department of Physics, 
    Via Roma, 56 - 53100 Siena, Italy }

\maketitle\abstracts{
The TOTEM experiment at the LHC is dedicated to the measurement of the  
total $pp$ cross section and to the study of elastic scattering and of
diffractive dissociation processes. TOTEM is here 
presented with a general overview on the main features of its experimental 
apparatus and of its physics programme.}

\section{Introduction}\label{sec:intro}

In order to fulfil its physics programme the TOTEM experiment~\cite{Totem_TDR} has to 
cope with the challenge of triggering and recording events in the very forward region with 
a good acceptance for particles produced at very small angles with respect to the beam. 
The evaluation of the total cross section with an error down to 1$\div$2\,$\%$ will 
in particular require the simultaneous measurement of the elastic scattering at low values 
for the squared four-momentum transfer ($|t| \sim (p\theta)^2$) as well as of the inelastic 
interaction rate with losses reduced to few per-cents. 
This involves the detection of scattered protons at a location very close 
to the beam, together with efficient forward particle detection in inelastic events. 
A flexible trigger provided by its detectors will allow to take data under all 
LHC running scenarios.
The combination of the CMS~\cite{CMS} and TOTEM experiments, representing the largest 
acceptance detector 
ever built at a hadron collider, will also allow the study of a wide range of diffractive
processes with an unprecedented coverage in rapidity. For this purpose the 
TOTEM trigger and data acquisition (DAQ) systems are designed to be compatible with the CMS 
ones, in order to allow common data taking periods foreseen at a later stage~\cite{CMS_TOTEM_TDR}.  
In the following, after a general overview of the experimental apparatus, the main 
features of the TOTEM physics programme will be described.

\section{Detector Components}\label{sec:detec}
\begin{figure}[htb!]
\epsfig{figure=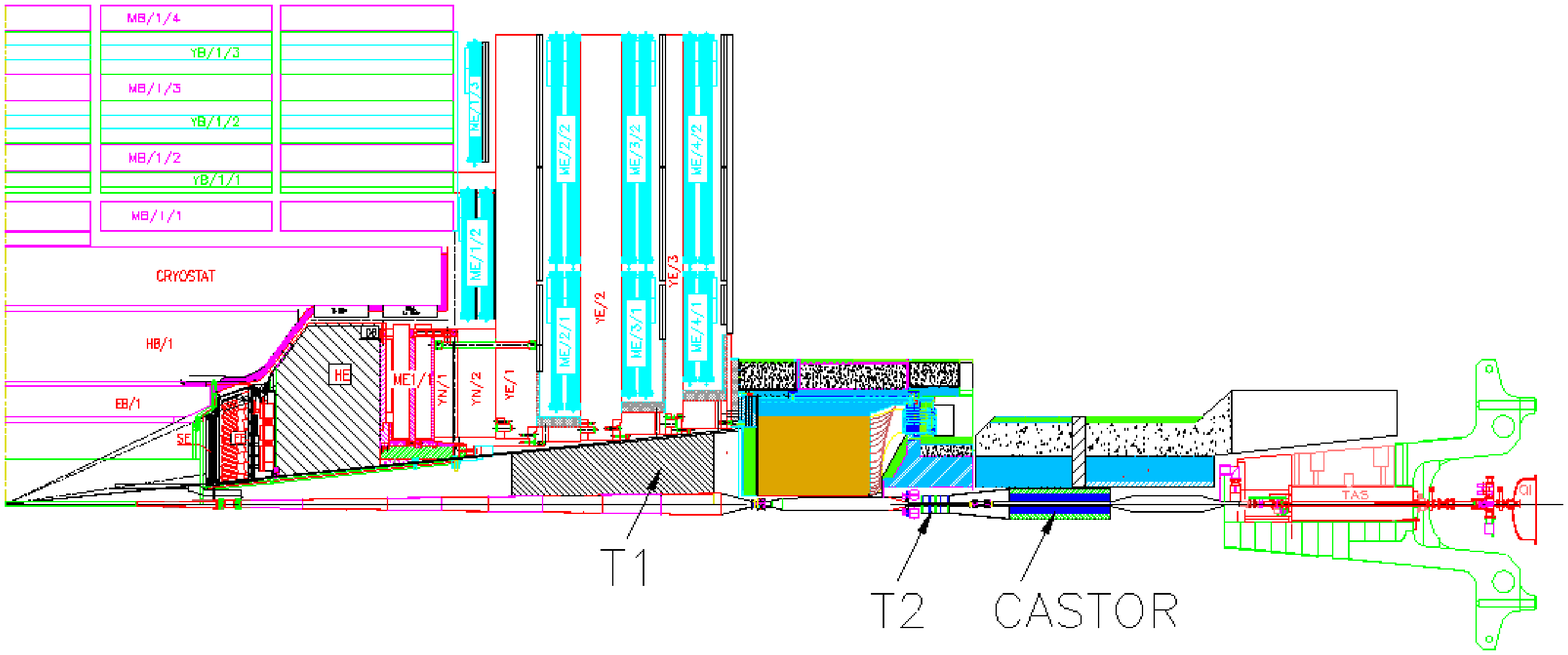,height=1.8in,width=10.0cm}
\vskip -4.5cm
\hskip 6.0cm
\epsfig{figure=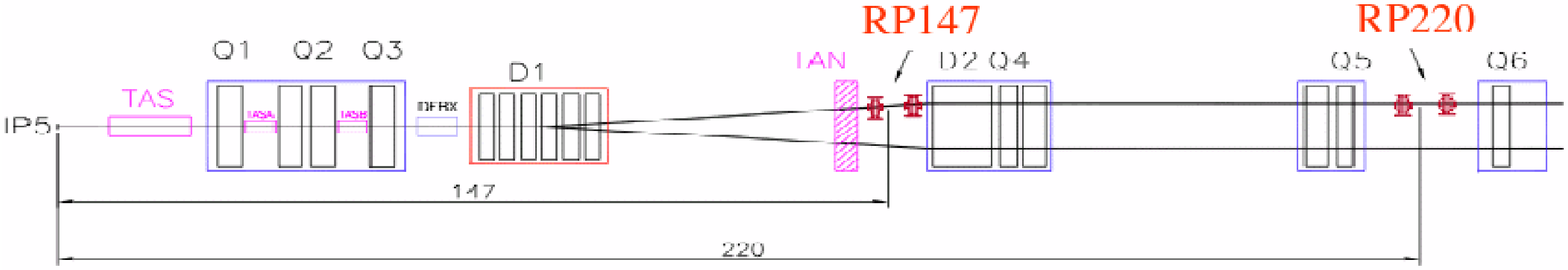,height=0.8in,width=10.0cm}
\vspace{1.5cm}
\caption{Right: Roman Pots location along the LHC beam-line. Left: T1/T2 location in the forward 
region of the CMS detector. All TOTEM detectors are located on both sides of IP5.}
\label{fig:totem_exp}
\end{figure}
Located on both sides of the interaction point IP5 (the same as for CMS) 
the TOTEM experimental apparatus comprises ``Roman Pots'' (RPs) detectors and the T1 and T2 
inelastic telescopes (Fig.~\ref{fig:totem_exp}).
The RPs, placed on the beam-pipe of the outgoing beam in two stations at about 147 m and 220 
m from IP5, are special movable beam-pipe insertions designed 
to detect ``leading'' protons with a scattering angle down to few $\mu$rad. 
T1 and T2, embedded inside the forward region of CMS, provide charged track reconstruction 
for 3.1 $<$ $|\eta|$ $<$ 6.5 ($\eta = -ln(tan\frac{\theta}{2})$) with a 2$\pi$ coverage and 
with a very good efficiency. These detectors, providing trigger signals with acceptance 
grater than 95$\%$ for all inelastic events, will allow the measurement of inelastic 
rates with small losses and will be also used for the reconstruction of the event 
interaction vertex, so to reject background events~\cite{CMS_TOTEM_TDR}. 
The read-out of all TOTEM sub-detectors is based on the digital VFAT chip~\cite{Totem_JNST}, 
specifically designed for TOTEM and characterized by trigger capabilities. 

\begin{figure}[htb!]
\psfig{figure=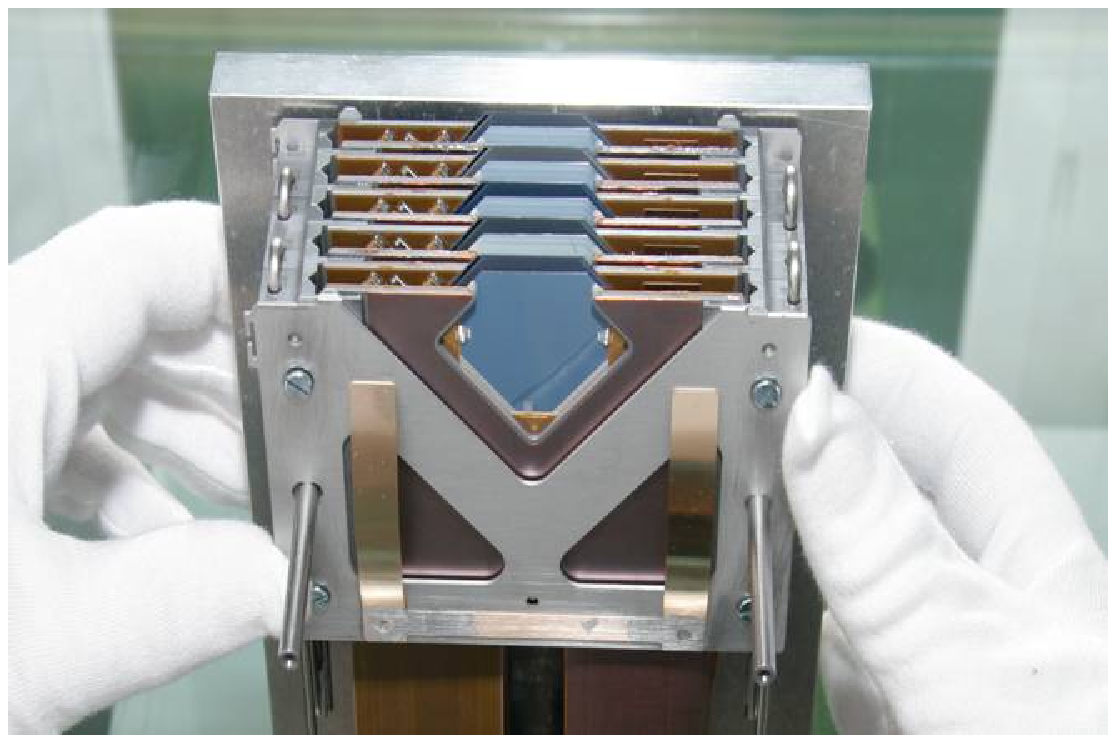,height=1.5in,width=0.32\linewidth}
\epsfig{figure=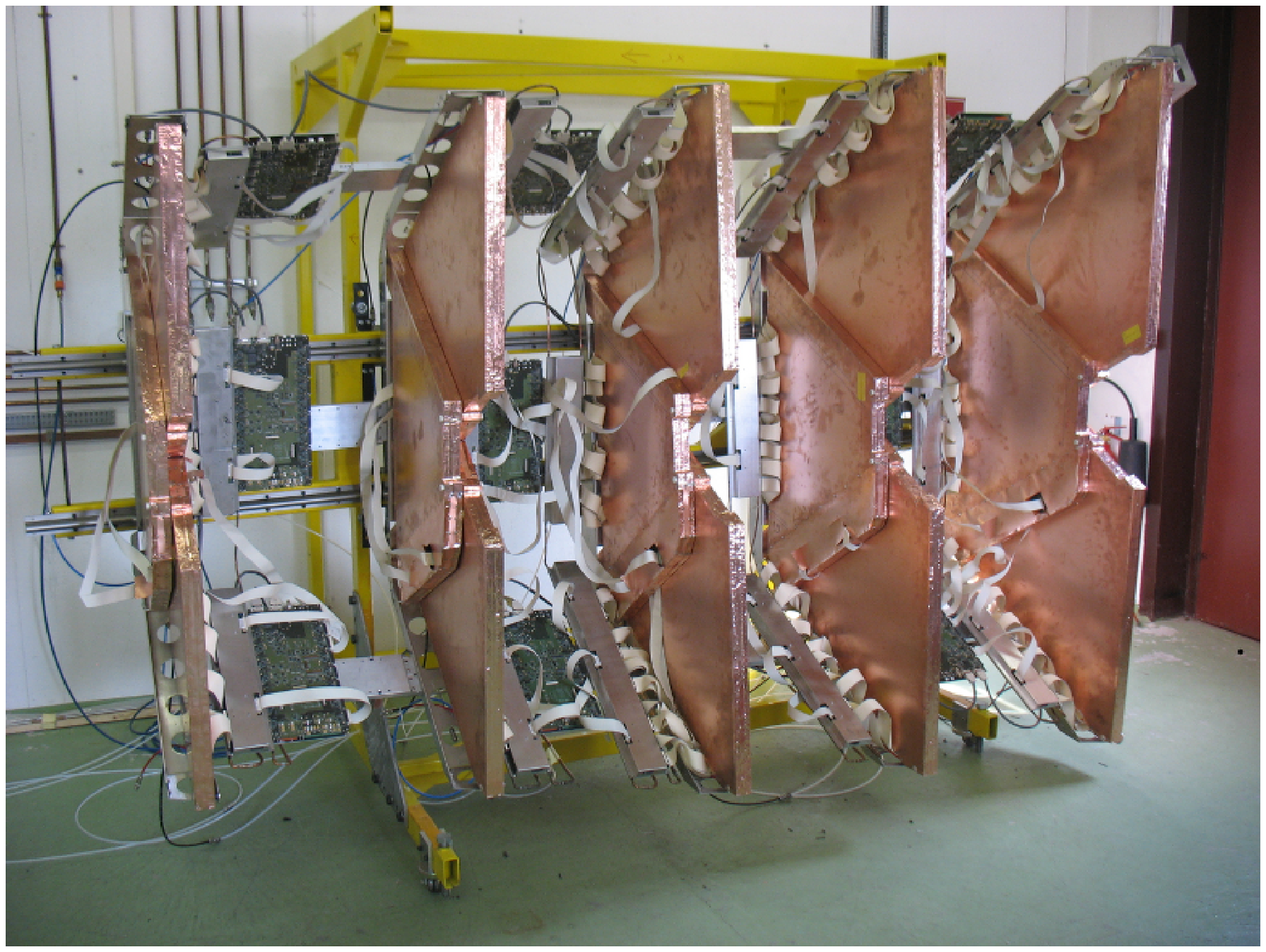,height=1.5in,width=0.32\linewidth }
\epsfig{figure=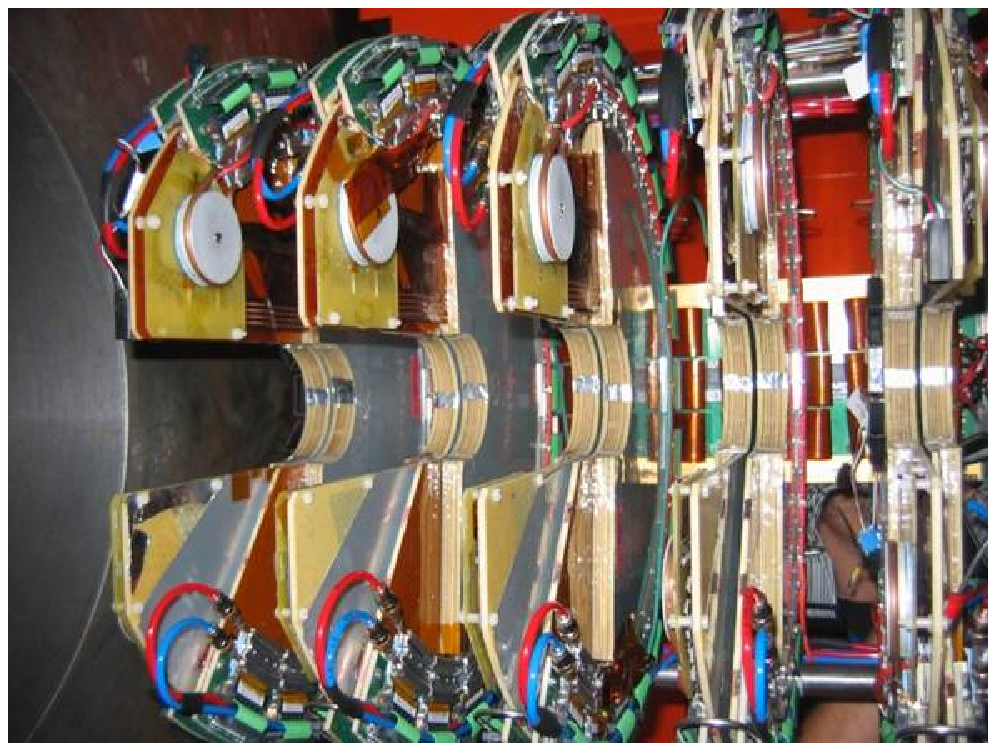,height=1.5in,width=0.32\linewidth}
\vspace{-0.2cm}
\caption{Left: silicon detectors hosted in one pot. Centre (right): 
            one half-arm of the T1 (T2) telescope.}
\label{fig:totem_det}
\end{figure}

The RPs host silicon detectors which are moved very close to the beam when it is 
in stable conditions. 
Each RP station is composed of two units in order to have 
a lever harm for local track reconstruction and trigger selections by track angle. 
Each unit consists of three pots, two vertical and one horizontal completing the 
acceptance for diffractively scattered protons. 
Each pot contains a stack of 10 planes of silicon strip detectors 
(Fig.~\ref{fig:totem_det}, left). Each plane has 512 strips (pitch of 
66 $\mu$m) allowing a single hit resolution of about 20 $\mu$m. 
As the detection of protons elastically scattered at angles down to few 
$\mu$rads requires a detector active area as close to the beam as $\sim$ 1 mm, 
a novel ``edgeless planar silicon'' detector technology has been developed 
for TOTEM RPs in order to have an edge dead zone minimized to only about 
50 $\mu$m~\cite{RP_Silicon}. 

Each T1 telescope arm, covering the range 3.1 $<$ $|\eta|$ $<$ 4.7 and located 
at $\sim$ 9 m from IP5, consists of five planes 
formed by six trapezoidal ``Cathode Strip Chambers'' (CSC)~\cite{Totem_TDR} 
(Fig.~\ref{fig:totem_det}, centre). These CSCs, with 10 mm thick gas gap and a gas mixture 
of Ar/CO$_2$/CF$_4$ ($40\%/50\%/10\%$), give three measurements of the charged particle 
coordinates with a spatial resolution of about 1 mm: anode wires (pitch of 3 mm), also giving 
level-1 trigger information, are parallel to the trapezoid base; 
cathode strips (pitch of 5 mm) are rotated by $\pm$ $60^o$ with respect to the wires.

The T2 telescope, based on ``Gas Electron Multiplier'' (GEM) technology~\cite{GEM}, 
extends charged track reconstruction to the range 5.3 $<$ $|\eta|$ $<$ 6.5~\cite{Totem_TDR}. 
Ten aligned detectors planes, having an almost semicircular shape, are combined to form one 
of the half-arms located at $\sim$ 13.5 m from IP5 (Fig.~\ref{fig:totem_det}, right). 
This novel gas detector technology is optimal for the T2 telescope 
thanks to its good spatial resolution, excellent rate capability and good 
resistance to radiation. The T2 GEMs are characterized by 
a triple-GEM structure and a gas mixture of Ar/CO$_2$ (70$\%$/30$\%$)~\cite{Totem_TDR}. 
The read-out board has two separate layers with different patterns: one with 256x2 concentric 
circular strips (80\,$\mu$m wide, pitch of 400\,$\mu$m), allowing track radial coordinate
reconstruction with a resolution of about 100\,$\mu$m; the other with a matrix of 24x65 pads 
(from 2x2\,mm$^2$ to 7x7\,mm$^2$ in size) providing level-1 trigger information and 
track azimuthal coordinate reconstruction.

\section{Physics Programme}\label{sec:phys}

The physics goals of the TOTEM experiment are the measurement of the total $pp$ cross section 
($\sigma_{tot}$) with a precision down to 1$\div$2\,$\%$, the study of the nuclear elastic 
$pp$ differential cross section ($d{\sigma}_{el} / dt$) over a wide range of 
$|t|$ ($\sim 10^{-3} < |t| < 10\,{\rm GeV}^{2}$) and the development of a comprehensive 
physics programme on diffractive dissociation, partially in cooperation with 
CMS. These studies will allow to distinguish among different models of soft proton 
interactions, giving a deeper understanding of proton structure. Furthermore, the study of 
charged particle flow in the forward region will give precious input to cosmic ray 
MonteCarlo simulations.  
\begin{figure}[htb!]
\epsfig{figure=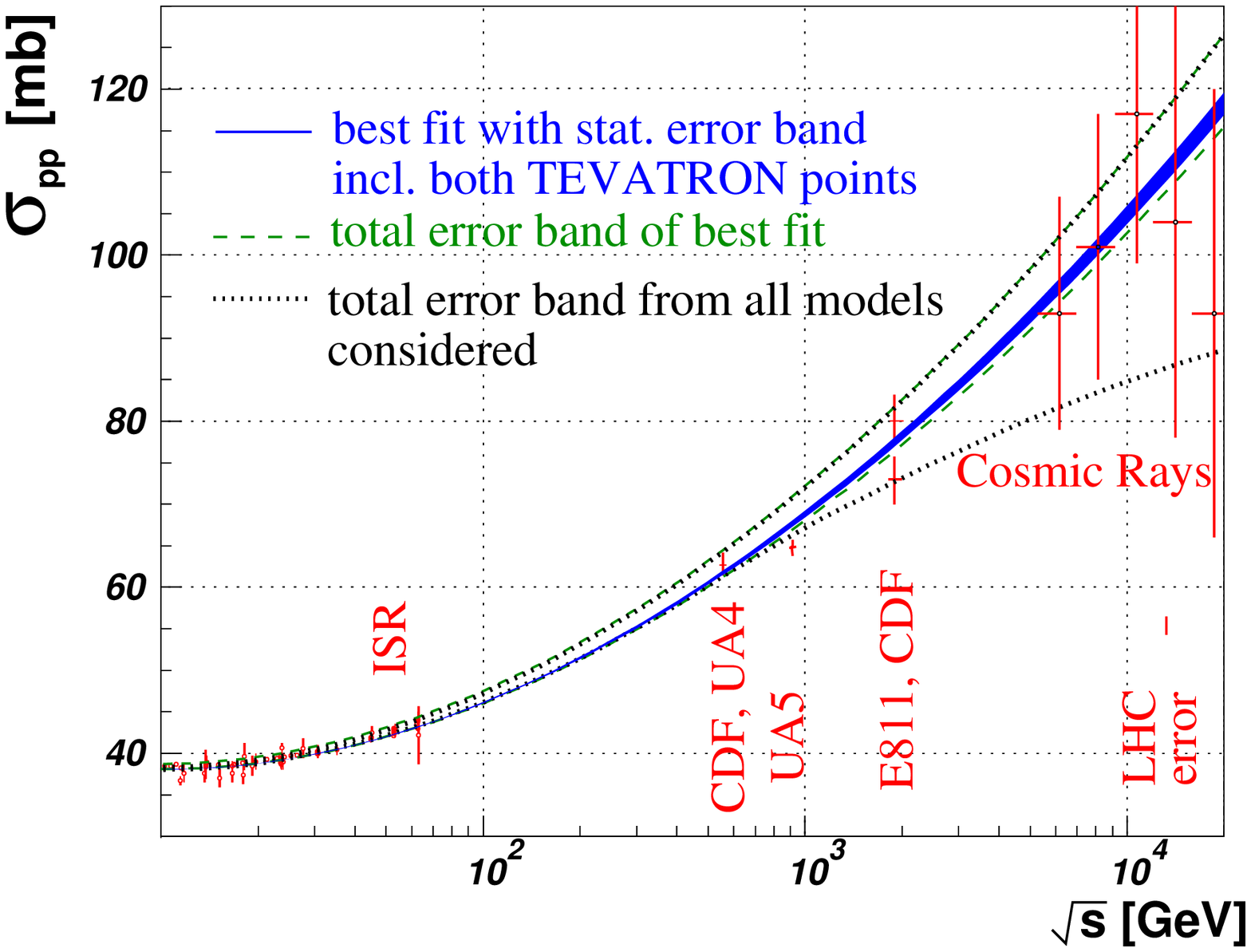,height=2.4in}
\epsfig{figure=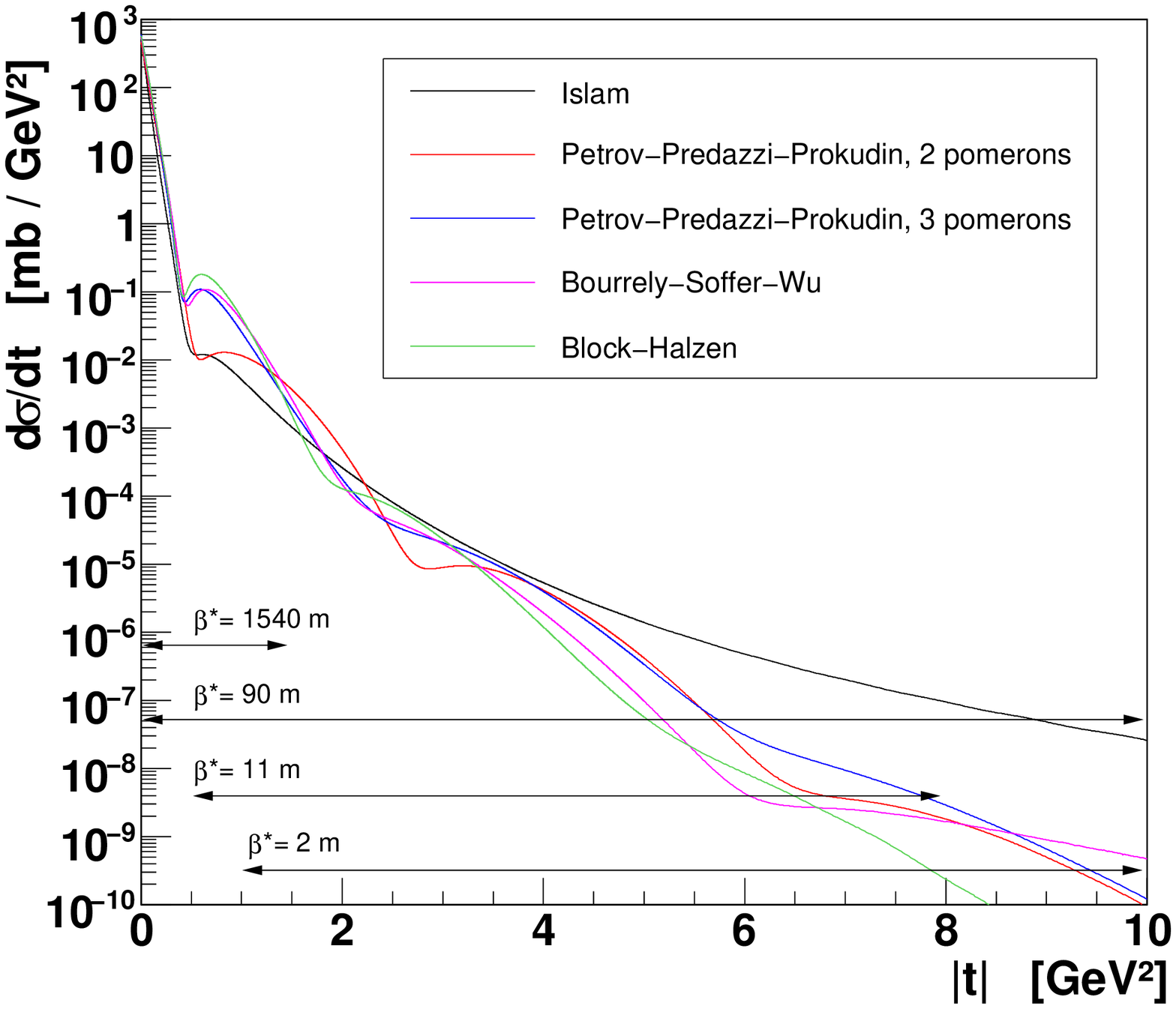,height=2.43in}
\vspace{-0.2cm}
\caption{Left: Fits from the COMPETE Coll. to available $pp$ and $p\bar p$ scattering 
and cosmic ray data. Right: $d{\sigma}_{el} / dt$ at LHC as predicted by 
different models; t-acceptance ranges for different machine optics are also shown.}
\label{fig:sigma_pp_el}
\end{figure}

Fig.~\ref{fig:sigma_pp_el} (left), summarizing the existing measurements
on $pp$ and $p\bar p$ scattering, shows predictions for $\sigma_{tot}$ from the COMPETE 
Collaboration based on fits according to different models~\cite{COMPETE}. 
The best fit predicts $\sigma_{tot} = 111.5 \pm 1.2 ^{+4.1}_{-2.1}$ mb for the LHC energy 
($\sqrt{s}$ = 14 TeV). The large uncertainties on available high energy data give a 
big error (90$\div$130 mb), depending on the model used for the extrapolation. 
TOTEM aims to measure $\sigma_{tot}$ with a precision down to 1$\div$2\,$\%$, 
therefore allowing to discriminate among the different models. The measurement will 
be based on the ``luminosity independent method'' which, combining the optical 
theorem with the total rate, gives $\sigma_{tot}$ and the machine 
luminosity ($\mathcal{L}$) as functions of measurable rates: 
\begin{equation}
   \sigma_{tot} = \frac{16 \pi}{1 + \rho^{2}} \cdot
   \frac{dN_{el}/dt |_{t=0}}{N_{el} + N_{inel}}
 \hskip 2.0cm
   \mathcal{L} = \frac{1 + \rho^{2}}{16 \pi} \cdot
   \frac{(N_{el} + N_{inel})^{2}}{dN_{el}/dt |_{t=0}} 
  \label{eq:stot_lumi}
\end{equation}
where $N_{el}$ and $N_{inel}$ are respectively the elastic and inelastic 
rate and $\rho$ is the ratio of real to imaginary part of the forward 
nuclear elastic scattering amplitude (given by theoretical predictions). 
Therefore, TOTEM will also provide an absolute measurement of $\mathcal{L}$ 
to be used for calibration purposes.
The uncertainty on the extrapolation of $dN_{el}/dt$ to $t = 0$ (optical point) depends on 
the acceptance for protons scattered at small $|t|$ values, hence 
at small angles. This requires a small beam angular divergence at the IP, which can be 
achieved in special runs with high $\beta^*$ machine optics and typically low $\mathcal{L}$. 
An approved optics with $\beta^* =$ 1540 m (and $\mathcal{L} \sim$ 10$^{28}$ cm$^{-2}$s$^{-1}$) 
will give a $\sigma_{tot}$ ($\mathcal{L}$) measurement at the level of 1$\div$2\,$\%$ 
(2\%). Another approved optics with $\beta^* =$ 90 m (and $\mathcal{L} \sim$
10$^{30}$ cm$^{-2}$s$^{-1}$), achievable without modifying the standard LHC injection optics, 
is expected to allow a preliminary $\sigma_{tot}$ ($\mathcal{L}$) measurement at the level 
of $\sim$ 5\% (7\%) in the first period of the LHC running. 
The experimental systematic error for 
the measurement with $\beta^* =$ 90 m will be dominated by the evaluation 
of $dN_{el}/dt |_{t=0}$, while with $\beta^{*} = 1540\,$m it will be 
dominated by the uncertainty on the corrections to trigger losses in Single 
Diffraction events for masses below $\sim$ 10\,GeV/c$^2$ ~\cite{Totem_JNST}. 
Given the high rates involved, the statistical error on $\sigma_{tot}$
will be negligible after few hours of data taking even at low $\mathcal{L}$.
The theoretical uncertainty related to the estimate of the $\rho$ parameter is expected 
to give a contribution on the relative uncertainty of less than 1.2\% (considering for 
instance the full error band on $\rho$ extrapolation as derived in ref~\cite{COMPETE}).
 
Fig.~\ref{fig:sigma_pp_el} (right) shows the distributions of $d{\sigma}_{el} / dt$
at $\sqrt{s}$ = 14 TeV as predicted by different models in the whole $|t|$-range 
accessible by TOTEM according to the different LHC optics settings~\cite{Totem_JNST}. 
Several regions are found at increasing $|t|$: for $|t| < 6.5 \times 10^{-4}$ GeV$^2$ 
(the Coulomb region)  
$d{\sigma}_{el} / dt \sim 1 / |t|^{2}$ is dominated by photon exchange;
for $|t|$ up to $\sim 10^{-3}$ GeV$^2$, the hadronic and Coulomb scattering 
interfere; for $10^{-3} < |t| < 0.5$ GeV$^2$ there is the hadronic region, 
e.g. described by ``single-Pomeron exchange'', characterized by an approximately 
exponential fall ($d{\sigma}_{el} / dt \sim {\rm e}^{-B\,|t|}$); 
the diffractive structure of the proton is then expected for $0.5 < |t| < 1$ GeV$^2$; 
for $|t| >1$ GeV$^2$ elastic collisions are described by pQCD, 
e.g. in terms of triple-gluon exchange with $d{\sigma}_{el} / dt \sim |t|^{-8}$. 
TOTEM will allow to discriminate among different models with a precise
measurement of $d{\sigma}_{el} / dt$ over all the accessible $t$-region, where 
$d{\sigma}_{el} / dt$ spans for over 11 orders of magnitude.
In the hadronic region, important for the extrapolation of $d{\sigma}_{el} / dt$ to $t=0$,  
a fit on $B(|t|)$ is typically performed in the $|t|_{min} < |t| < 0.25$ GeV$^2$ range,  
$|t|_{min}$ depending on the acceptance for protons scattered at small angles, hence on 
the beam optics. 

Diffractive (due to colour singlet exchange) and non-diffractive 
(due to colour exchange) inelastic interactions represent a big 
fraction (around $60\div 70$ $\%$) of $\sigma_{tot}$.
Nevertheless many details of these processes, with close ties to proton structure 
and low-energy QCD, are still poorly understood. 
The majority of diffractive events exhibits intact (``leading'') protons 
characterized by their $t$ and fractional momentum loss $\xi \equiv \Delta p/p$. 
TOTEM will be able to measure $\xi$-, $t$- and mass-distributions with acceptances 
depending on the beam optics. It will also study the charged particle flow in 
the forward region, providing a significant contribution to the 
understanding of cosmic ray physics. The existing models  
give in fact predictions on energy flow, multiplicity and other quantities related 
to cosmic ray air showers, with significant inconsistencies in the forward region.
Furthermore, the integration of TOTEM with the CMS detector will offer 
the possibility of more detailed studies on inelastic events, including hard 
diffraction~\cite{CMS_TOTEM_TDR}. 

\section{Summary and Conclusions}

The TOTEM experiment will be ready for data taking at the LHC start. 
Running under all beam conditions, it will perform an important 
and exciting physics programme involving the measurement of $\sigma_{tot}$ and 
$d{\sigma}_{el} / dt$ in $pp$ interactions as well as studies on diffractive 
precesses and on forward charged particle production. 
Special high ${\beta}^*$ runs will be required in order to measure 
$\sigma_{tot}$ at the level of $\sim$ 5 $\%$ (early measurement 
with ${\beta}^*$ = 90 m) and $\sim 1\div 2$ $\%$ (${\beta}^*$ = 1540 m). 
$d{\sigma}_{el} / dt$ will be studied in the range 
$\sim 10^{-3} < |t| < 10\,{\rm GeV}^{2}$ allowing to distinguish 
among several theoretical predictions. A common physics programme 
with CMS on soft and hard diffraction as well as on forward particle 
production studies is also foreseen in a later stage.

\end{document}